\documentclass[a4paper, 10.5pt]{article}
\usepackage{graphicx}
\usepackage{tabularx}
\begin{document}

\begin{center}
{\Large Effects of multiple pairs on visibility measurements of entangled photons generated by spontaneous parametric processes} 
\end{center}

\begin{center}
\today\\
Hiroki Takesue and Kaoru Shimizu\\
NTT Basic Research Laboratories, NTT Corporation\\
3-1 Morinosato Wakamiya, Atsugi, Kanagawa 243-0198, Japan\\
and CREST, Japan Science and Technology Agency\\
4-1-8 Honcho, Kawaguchi, Saitama 332-0012, Japan
\end{center}

\begin{center}
Abstract
\end{center}

Entangled photon-pair sources based on spontaneous parametric processes are widely used in photonic quantum information experiments. 
In this paper, we clarify the relationship between average photon-pair number and the visibility of two-photon interference (TPI) using those entanglement sources.
We consider sources that generate distinguishable and indistinguishable entangled photon pairs, assuming coincidence measurements that use threshold detectors. 
We present formulas for the TPI visibility of a polarization entanglement that take account of all the high-order multi-pair emission events. Moreover, we show that the formulas can be approximated with simple functions of the average pair number when the photon collection efficiency is small. As a result, we reveal that an indistinguishable entangled pair provides better visibility than a distinguishable one.

\newpage
\section{Introduction} \label{intro}

Entanglement is an important resource in quantum information technologies, such as quantum key distribution (QKD) and quantum computation \cite{pqi}. An entangled state of photons are especially important for distributing entanglement over long distance. The most common way of generating entangled photon pairs is to use spontaneous parametric processes, including spontaneous parametric downconversion (SPDC) \cite{kwiat1,kwiat2,brendel,yoshi,takesueppln} and spontaneous four-wave mixing (SFWM) \cite{takesue1,li,takesue3,rarity,fan}. 
These processes have been used for generating entangled states in polarization, energy-time, time-bin, and spatial modes. 
These entangled photon-pair sources have been used for a number of photonic quantum information experiments, such as the violation of Bell's inequality \cite{kwiat1,kwiat2,takesue1,li}, quantum teleportation \cite{qt}, entanglement swapping \cite{pan}, QKD \cite{qkd1,qkd2}, and quantum logic gates \cite{kok}. In addition, entanglement generation in the 1.5-$\mu$m telecom band is now being intensively studied with a view to realizing quantum communication systems over optical fiber.

In previous experiments, two-photon interference (TPI) is often used to confirm the generation of entanglement. TPI visibility is a good measure of entanglement quality. 
It is well known that TPI visibility is limited by ``accidental coincidences". Entangled photon-pair sources based on spontaneous parametric processes emit multi-pairs with finite probabilities, and these multi-pair events cause coincidences with uncorrelated photons, which are called accidental coincidences. We cannot suppress the degradation in visibility caused by accidental coincidences completely unless we use photon number resolving single photon detectors with perfect photon collection efficiencies. In fact, most previous experiments used ``threshold detectors", which are single photon detectors without a photon number resolution ability, and the photon collection efficiencies were much smaller than unity. Therefore, it is essential to take account of accidental coincidences when evaluating entangled photon pair sources. 

Since accidental coincidences are caused by multi-pair events, it is obvious that the probability of observing an accidental coincidence depends on the number distribution of the entangled photon pairs. 
The work by Luo and Chan \cite{luo} and Lim et al. \cite{lim} suggests that the pair number distribution becomes thermal-like when the photons generated in multi-pair emission events are quantum-mechanically indistinguishable. On the other hand, it is known that the distribution becomes Poissonian when the photons are distinguishable \cite{longd,kuzuku}. 
Even when pairs are generated in a single spatial mode, the actual distribution changes from a thermal-like to a Poisson distribution depending on the conditions of the pump pulse width, the length of nonlinear interaction, and the coherence time of the pairs. 
In any case, the visibilities improve when the average pair number is reduced by using a weaker pump. 
Therefore, many previous experiments used a very weak pump so that the accidental coincidences caused by multi-pair emission became negligible. 
However, in certain quantum information systems, it is sometimes better to use a relatively large average pair number, accepting the relatively large number of accidental coincidences. 
For example, in an entanglement-based QKD system, we can optimize the average pair number to maximize the secure key rate: i.e. we can often increase the secure key rate by using a relatively large average pair number while accepting an increased error rate.  
Since the error rate is directly linked to the TPI visibility, it is meaningful to know the relationship between the average pair number and the visibility.

The purpose of this study is to clarify the relationship between the average pair number and the quality of the entangled photon pairs generated by spontaneous parametric processes.  
Although the TPI visibility of distinguishable entanglement sources is considered briefly in \cite{longd,kuzuku}, its full derivation has not yet been reported. Moreover, the TPI visibility of an indistinguishable entanglement source has not yet been clarified. Therefore in this paper, we present the formulas for the TPI visibilities of distinguishable and indistinguishable entangled photon-pair sources. We assume the use of threshold detectors,  since most of the entanglement generation experiments conducted so far have used this type of detector. 
In particular, when the collection efficiency of the photon is small (this assumption can be validated for most of the previous experimental conditions), the TPI visibilities are given by simple functions of the average pair number. 
Then, the density matrices and concurrences of distinguishable and indistinguishable entanglement sources are given as a function of the average pair number. 
We also present formulas modified for time-bin entanglement measurements. 
We believe that our results will be useful for designing and evaluating quantum information experiments using entangled photon-pair sources based on spontaneous parametric processes.

The organization of this paper is as follows. 
In section \ref{process}, we review the pair number distribution of indistinguishable and distinguishable entangled photon pairs, which become thermal-like and Poissonian, respectively. 
In section \ref{tpi}, the TPI visibilities of two types of entanglement sources are given as a function of average pair number and collection efficiency. 
Section \ref{qst} shows that the density matrices of those sources can be theoretically obtained if the average pair number and the collection efficiencies are known. Here, we also calculate the theoretical concurrence using the obtained density matrices. 
Section \ref{dc} considers correction terms that take account of the detector dark counts. 
Section \ref{tb} describes formulas that are modified for time-bin entangled photon pairs. 
Finally, the obtained results are summarized in section \ref{summary}.

\section{Number distribution of entangled photon pairs generated by spontaneous parametric processes} \label{process}

In this section, we derive the pair number distributions when the generated multi-pairs are quantum-mechanically indistinguishable, and distinguishable.

\subsection{Indistinguishable entangled photon pairs}

When photon pairs are generated in the same mode by a spontaneous parametric process, the photons obtained in a multi-emission event are indistinguishable. 
According to \cite{dfw}, an interaction Hamiltonian for single-mode correlated photon pair generation by spontaneous parametric processes at the origin of the time evolution can be expressed as
\begin{equation}
\hat{H}_{int} = i \hbar \chi (a_s^\dagger a_i^\dagger - a_s a_i), \label{intpp}
\end{equation}
where subscripts $s$ and $i$ denote the signal and idler modes, respectively. 
$\chi$ in the above equation for SPDC and SFWM is expressed as
\begin{eqnarray}
\chi_{SPDC} &=& c \chi^{(2)} a_p, \\
\chi_{SFWM} &=& c' \chi^{(3)} a_{p1} a_{p2},
\end{eqnarray}
where $a_p$, $a_{p1}$ and $a_{p2}$ denote pump field amplitudes. Here, we assume that the pumps are strong and can be treated as classical oscillators. 

Entangled photon pairs are generated when two spontaneous parametric processes in different modes occur coherently. Here we assume polarization entangled photon pairs, which are generated by the following interaction Hamiltonian.
\begin{equation}
\hat{H}_{intE} = \hat{H}_{intH} + \hat{H}_{intV}. \label{inthv}
\end{equation}
Here, {\it H} and {\it V} denote the horizontal and vertical polarization modes, respectively, and $\hat{H}_{intH}$ and $\hat{H}_{intV}$ are the interaction Hamiltonians for generating {\it H} and {\it V} polarization photon pairs given by
\begin{eqnarray}
\hat{H}_{intH} &=& i \hbar \chi (a_{sH}^\dagger a_{iH}^\dagger - a_{sH} a_{iH}), \\
\hat{H}_{intV} &=& i \hbar \chi (a_{sV}^\dagger a_{iV}^\dagger - a_{sV} a_{iV}). \end{eqnarray}
Using $\hat{H}_{intE}$, the time evolution of state is described with the following equation. 
\begin{equation}
|\Psi\rangle = \exp\left\{-i \frac{\hat{H}_{intE}}{\hbar} t  \right\} |0\rangle \label{ev1}
\end{equation}
According to the disentangling theorem shown as Eq. (5.63) in \cite{dfw}, the exponential part can be rewritten as
\begin{equation}
\exp\left\{-i \frac{\hat{H}_{intE}}{\hbar} t  \right\} = e^{\Gamma a_{sH}^\dagger a_{iH}^\dagger} e^{-g(a_{sH}^\dagger a_{sH} + a_{iH}^\dagger a_{iH}+1)} e^{-\Gamma a_{sH} a_{iH}} 
e^{\Gamma a_{sV}^\dagger a_{iV}^\dagger} e^{-g(a_{sV}^\dagger a_{sV} + a_{iV}^\dagger a_{iV}+1)} e^{-\Gamma a_{sV} a_{iV}},
\end{equation}
where 
\begin{eqnarray}
\Gamma &=& \tanh \chi t, \\
g &=& \ln (\cosh \chi t).
\end{eqnarray}
With this equation, and discarding the term that becomes zero, Eq. (\ref{ev1}) is expressed as
\begin{eqnarray}
|\Psi\rangle &=& \frac{1}{C^2} \exp (\hat{H}_2) |0\rangle \nonumber \\
&=& \frac{1}{C^2} \left\{ 1 + \hat{H}_2 + \frac{1}{2!} \hat{H}_2^2 + \cdots + \frac{1}{x!} \hat{H}_2^x + \cdots \right\} |0\rangle. \label{ev2}
\end{eqnarray}
Here, $\hat{H}_2$ is defined as 
\begin{equation}
\hat{H}_2 = \Gamma (a_{sH}^\dagger a_{iH}^\dagger + a_{sV}^\dagger a_{iV}^\dagger), 
\end{equation}
and $C=\cosh \chi t$. 

The first order term in Eq. (\ref{ev2}) gives an entangled photon pair state, which is calculated as
\begin{equation}
|\Psi_1\rangle = \frac{1}{C^2} \hat{H}_2 |0\rangle = \frac{\Gamma}{C^2} \left\{|1,0\rangle_s |1,0\rangle_i + |0,1\rangle_s |0,1\rangle_i \right\}, 
\end{equation}
where $|a,b\rangle_x$ denotes that in mode $x$, there are $a$ photons with H polarization and $b$ photons with V polarization. 

Similarly, the second order term, in which two-pair state is generated, is expressed as follows. 
\begin{equation}
|\Psi_{2indis}\rangle = \frac{1}{C^2} \hat{H}_2^2 |0\rangle = \frac{\Gamma^2}{C^2} \left\{|2,0\rangle_s |2,0\rangle_i + |1,1\rangle_s |1,1\rangle_i+|0,2\rangle_s |0,2\rangle_i \right\}
\end{equation}
This is the state that is described as an entangled two-photon polarization state in \cite{tsujino}. 

The $x$th order term can be expressed as 
\begin{equation}
|\Psi_{xindis}\rangle = \frac{\Gamma^x}{C^2} \sum_{k=0}^x  |x-k,k\rangle_s |x-k,k\rangle_i=\sqrt{x+1} \frac{\Gamma^x}{C^2} |\psi_x\rangle, \label{xorder}
\end{equation}
where $x$ is a positive integer and $|\psi_x\rangle$ is a normalized quantum state of $x$-pair given by
\begin{equation}
|\psi_x\rangle = \frac{1}{\sqrt{x+1}} \sum_{k=0}^x  |x-k,k\rangle_s |x-k,k\rangle_i. \label{indisxpair}
\end{equation}

Equation (\ref{xorder}) shows that the probability of generating $x$ pairs $P_{indis}(x)$ is given by
\begin{equation}
P_{indis}(x) = \frac{(x+1) \Gamma^{2x}}{C^4} = \frac{(x+1) \tanh^{2x} \chi t}{\cosh^4 \chi t}, \label{unnormthermal}
\end{equation}
and the average photon-pair number $\mu$ is given by
\begin{equation}
\mu = \sum_{x=0}^{\infty} x P_{indis}(x) = 2 \sinh^2 \chi t.
\end{equation}
When the nonlinear medium is pumped by pulses, the average pair number is defined as a number per pulse. When it is pumped by a continuous light, the average pair number is defined as a number per temporal resolution of a measurement system, which is often determined by the temporal resolution of the photon detectors.  

Using a relationship $\cosh^2 \chi t - \sinh^2 \chi t = 1$, the probability of generating $x$ pairs is expressed as a function of $\mu$, which is 
\begin{equation}
P_{indis} (\mu,x) = (1+x) \frac{\left(\frac{\mu}{2}\right)^x}{(1+\frac{\mu}{2})^{x+2}}. \label{thermal}
\end{equation}

\subsection{Distinguishable entangled photon pairs}

We then consider the situation in which a pump can generate entangled photon pairs in many different modes. 
This situation is typically observed when a continuous wave pump is used or the pump pulse width is much longer than the coherence time of the photon pairs \cite{tapster,reid}. In those cases, the generated pairs are distinguishable in the time domain. Also, even if the pump pulse width is much shorter than the photon pair coherent time, the group delay difference between pump pulse and photon pair in the nonlinear crystal can lead to temporal distinguishability.
In those cases, the pairs generated in multi-emission events are mutually distinguishable. 
For example, when two pairs of entangled photons are generated through independent spontaneous parametric processes in temporal modes 1 and 2, the joint state is given by
\begin{equation}
|\Psi_{2dis}\rangle = \frac{1}{2} \left\{|1,0\rangle_{s1} |1,0\rangle_{i1} + |0,1\rangle_{s1} |0,1\rangle_{i1} \right\} \otimes \left\{|1,0\rangle_{s2} |1,0\rangle_{i2} + |0,1\rangle_{s2} |0,1\rangle_{i2} \right\}. \label{2dis}
\end{equation}
In the following, we explain why the number distribution of distinguishable entangled photon pairs becomes Poissonian. 

Here we consider the temporal distinguishability that arises from the use of a broad pump pulse. 
We define an integer $N$ as follows. 
\begin{equation}
N = \frac{\Delta t_{pump}}{\Delta t_{pp}}
\end{equation}
Here, $\Delta t_{pump}$ and $\Delta t_{pp}$ denote the pump pulse width and coherence time of the photon pairs, respectively. 
The physical meaning of $N$ is the number of different temporal modes excited by the pump pulse. 
Let us consider a case where $x$ pairs are created. If $N$ is far larger than $x$, the probability that two or more pairs are in the same temporal mode 
is negligible. 
In this case, the $x$ pairs are all distinguishable. This means that the probability of generating $x$ pairs is simply given by multiplying of the single pair generation probability, which is given by Eq. (\ref{unnormthermal}) with $x=1$.  
The number of combinations for choosing the temporal modes of the $x$ pairs in $N$ modes is given by $_N C_x$. 
Therefore, the probability of generating $x$ pairs is expressed as\begin{equation}
P_p (N,x) \simeq  a ._N C_x \left(\frac{2 \Gamma^{2}}{C^4}\right)^x \simeq a \left(\frac{2 \Gamma^{2} N}{C^4}\right)^x \frac{1}{x!},
\end{equation} 
where $a$ is a constant for normalization, and an approximation $_N C_x \simeq N^x/x!$ is used. By setting $\left(\frac{2 \Gamma^{2} N}{C^4}\right)=\mu$ and $a=e^{-\mu}$, the probability of generating $x$ distinguishable pairs is obtained as
\begin{equation}
P_p (\mu,x) \simeq e^{-\mu} \frac{\mu^x}{x!}. \label{poi}
\end{equation}
Thus, when $N>>x$, the pair number distribution becomes Poissonian. It is obvious that $\mu$ gives the average pair number.


\section{Two-photon interference measurement} \label{tpi} 

\begin{figure}[htb]

\centerline{\includegraphics[width=.9\linewidth]{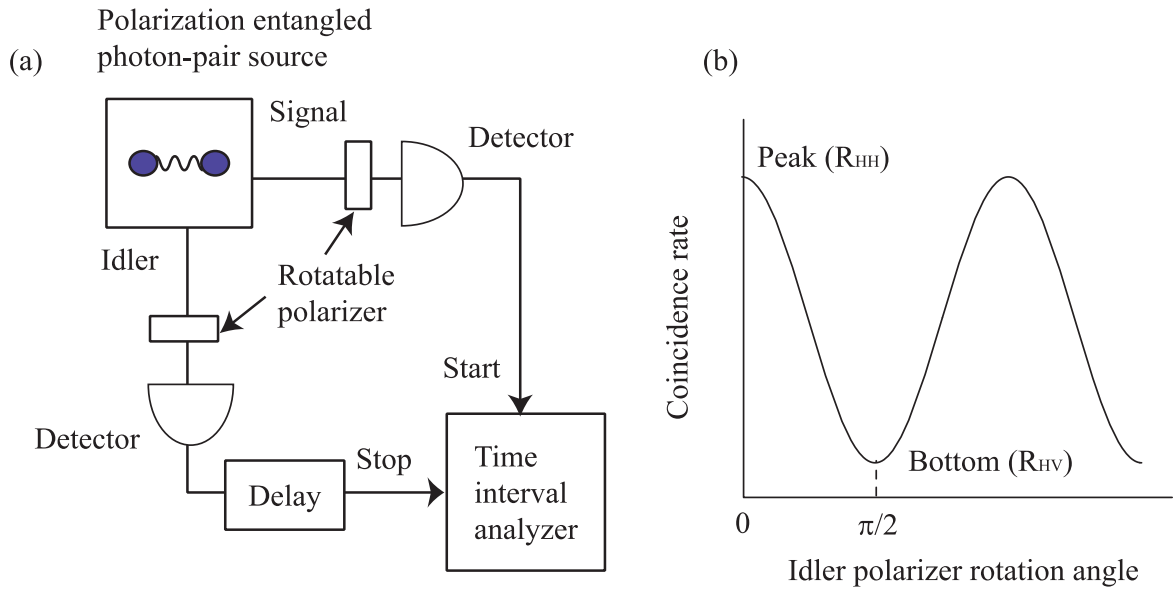}}

\caption{(a) TPI visibility measurement setup, (b) typical TPI waveform. } \label{vsetup}
\end{figure}

In this section, we consider the visibility of the TPI measurement. We assume polarization entangled photon-pair states whose first-order approximations are given by
\begin{equation}
|\Phi^+\rangle = \frac{1}{\sqrt{2}}\left(|1,0\rangle_s |1,0\rangle_i + |0,1\rangle_s |0,1\rangle_i \right). \label{ent}
\end{equation}
In the following, we analyze the way in which the multi-pair emission degrades the visibilities of TPI fringes for distinguishable and indistinguishable entangled photon pairs.  In this analysis we discard all the other causes of visibility degradation such as a deviation of the polarization reference between the signal and idler measurements and the dark counts of the detectors.

We assume the experimental setup shown in Fig. \ref{vsetup} (a). A polarization entangled photon-pair source generates a signal and an idler photon. Each photon is input into a rotatable polarizer that is followed by a threshold detector. When $x$ photons are incident on a threshold detector with a collection efficiency of $\alpha$, the detection probability is obtained as
$
1-(1-\alpha)^x
$.
Here, the collection efficiency is a value that includes all the optical losses and the quantum efficiency of the detector.  
The detection signals from the detectors are input into a time interval analyzer (TIA) to measure the coincidence counts. We assume that the polarizer for the signal is fixed at {\it H} polarization, and then the coincidence rate is measured as a function of idler polarizer angle. 
Figure \ref{vsetup} (b) shows a typical TPI fringe for the (approximated) state given by Eq. (\ref{ent}), where the coincidence rate is illustrated as a function of idler polarizer angle from {\it H} polarization. When we set the idler polarizer angle at {\it H} and {\it V} polarizations, we observe the``peak" and ``bottom" of the TPI fringe, respectively. We denote the coincidence rates at the peak and bottom as $R_{HH}$ and $R_{HV}$, respectively. Then, the visibility of the waveform is obtained as 
\begin{equation}
v=\frac{R_{HH}-R_{HV}}{R_{HH}+R_{HV}}. \label{visibility}
\end{equation}
We also introduce a contrast parameter $\tau$ defined as
\begin{equation}
\tau=\frac{R_{HH}}{R_{HV}} \label{contrast}
\end{equation}
In the following, we theoretically calculate $R_{HH}$ and $R_{HV}$ to obtain the $v$ and $\tau$ values for the distinguishable and indistinguishable entangled photon pairs.

\subsection{Distinguishable entangled photon pairs}

Here, we consider the TPI visibility for the distinguishable entangled pairs whose pair number distribution is given by Eq. (\ref{poi}).

\subsubsection{Coincidence rate at fringe peak}

We assume that both the signal and idler polarizers are set at {\it H} polarization to obtain coincidence rate $R_{HH}$. 

First, let us consider a case in which a single pair is created with a probability of $P_p (\mu,1)$ given by Eq. (\ref{poi}). The patterns of polarization for a photon pair are only both {\it H} or both {\it V}, which occurs with the same emergence probability of 1/2. When both photons have {\it H} polarization, a coincidence count is observed with a detection probability of $\alpha^2$. Therefore, the overall coincidence probability when one pair is created is given by $\frac{1}{2}\alpha^2$.

\begin{center}
Table 1: Single pair\\
\begin{tabularx}{.60\linewidth}{ c|c| c }
\hline
Pattern & Emergence probability & Detection probability  
  \\
\hline
 $HH$ & $\frac{1}{2}$ & $\alpha^2$   \\
\hline
$VV$ & $\frac{1}{2}$ &0   \\
\hline
\end{tabularx}
\end{center}

Then, we consider a case where two distinguishable pairs are created. The state is given by Eq. (\ref{2dis}). 
The two pairs are created with a probability of $P_p (\mu,2)$.
Here we should note that the two pairs are distinguishable, so we denote them pair 1 and pair 2. The patterns of polarization, the emergence probabilities of the corresponding patterns, and the probabilities of detecting coincidences for the corresponding patterns using threshold detectors are summarized in Table 2. The table indicates that the total coincidence probability of two pairs is given by $
\frac{1}{4} \{1-(1-\alpha)^2\}^2 + \frac{1}{2} \alpha^2
$. 
 
\begin{center}

Table 2: Two pairs \\

\begin{tabularx}{.60\linewidth}{ c|c| c }
\hline
Pattern & Emergence probability & Detection probability  
  \\
\hline
 $\begin{array}{c} HH \\ HH \end{array}$ & $\frac{1}{4}$ & $\{1-(1-\alpha)^2\}^2$   \\
\hline
 $\begin{array}{c} HH \\ VV \end{array}$ & $\frac{1}{4}$ & $\alpha^2$   \\
\hline
$\begin{array}{c} VV \\ HH \end{array}$ & $\frac{1}{4}$ & $\alpha^2$   \\
\hline
$\begin{array}{c} VV \\ VV \end{array}$ & $\frac{1}{4}$ & 0   \\
\hline
\end{tabularx}
\end{center}

Then let us consider the $x$-pair case, where $x$ is a positive integer. 
A distinguishable $x$-pair state is expressed as
\begin{equation}
|\Psi_{xdis}\rangle = \left(\frac{1}{\sqrt{2}}\right)^x \bigotimes_{k=1}^x \{|1,0\rangle_{sk} |1,0\rangle_{ik} + |0,1\rangle_{sk} |0,1\rangle_{ik} \}. \label{pxpair}
\end{equation}
The total number of patterns is given by $2^x$, and so the emergence probability of each pattern is given by $1/2^x$. 
If $y$ pairs are in the $VV$ state ($y$: nonnegative integer), the number of patterns of different sets of $HH$ and $VV$ pairs is expressed as $_xC_y=\frac{x!}{y!(x-y)!}$, with a detection probability of $\{1-(1-\alpha)^{x-y}\}^2$. Thus, the coincidence probability in the setup when $x$ pairs are generated is given by
\begin{equation}
f(x) = \sum_{y=0}^x \frac{1}{2^x} \frac{x!}{y!(x-y)!} \{1-(1-\alpha)^{x-y}\}^2
\end{equation}

Since the probability of observing $x$ photon pairs is given by $P_x (\mu)$, the overall coincidence rate is obtained with the following equation.
\begin{equation}
R_{HHp} = \sum_{x=0}^\infty f(x) P_p (\mu,x) \label{rhhpoi}
\end{equation}

When $\alpha << 1$, the result shown in Eq. (\ref{rhhpoi}) can be approximated as follows. 
Using $\{1-(1-\alpha)^{x-y}\}^2 \simeq (x-y)^2 \alpha^2$,
\begin{eqnarray}
R_{HHp} &\simeq& \alpha^2 \sum_{x=0}^\infty \left\{
\sum_{y=0}^x \frac{1}{2^x} \frac{x!}{y! (x-y)!} (x-y)^2 \frac{\mu^x}{x!} e^{-\mu}  \right\} \nonumber \\
&=& \alpha^2 \left( \frac{\mu}{2} + \frac{\mu^2}{4}\right). \label{approxhhpoi}
\end{eqnarray}
The detailed derivation of the above equation is shown with Eqs. (\ref{temp}) to (\ref{apphhp}) in Appendix \ref{appcal}.

\subsubsection{Coincidence rate at fringe bottom}

Here we assume that the signal and the idler polarizer are set at {\it H} and {\it V} polarizations, respectively, to obtain the coincidence rate $R_{HV}$.  
As with the ``peak" case, when $x$ pairs are created, the total number of patterns is given by $2^x$, and the emergence probability of each pattern is given by $1/2^x$. 
If $y$ pairs are in the $VV$ state, the number of patterns of different sets of $HH$ and $VV$ pairs is expressed as $_xC_y=\frac{x!}{y!(x-y)!}$. Since the numbers of {\it VV} and {\it HH} pairs are $y$ and $x-y$, respectively, the detection probability is given by $\{1-(1-\alpha)^{x-y}\}\{1-(1-\alpha)^{y}\}$. Thus, the coincidence detection probability in the setup when $x$ pairs are generated is given by
\begin{equation}
g(x) = \sum_{y=0}^x \frac{1}{2^x} \frac{x!}{y!(x-y)!} \{1-(1-\alpha)^{x-y}\}\{1-(1-\alpha)^{y}\}
\end{equation}
The overall coincidence rate is obtained with the following equation.
\begin{equation}
R_{HVp} = \sum_{x=0}^\infty g(x) P_p (\mu,x) \label{rhvpoi}
\end{equation}

As with the peak coincidence rate, we can simplify the above equation when $\alpha <<1$. 
Using $\{1-(1-\alpha)^{x-y}\}\{1-(1-\alpha)^{y}\} \simeq (x-y)y\alpha^2$, $R_{HVp}$ is reduced to
\begin{equation}
R_{HVp} \simeq \sum_{x=0}^\infty \left\{ \sum_{y=0}^x \frac{1}{2^x} \frac{x!}{y!(x-y)!} (x-y) y \alpha^2 \frac{\mu^x}{x!} e^{-\mu} \right\} 
= \frac{\alpha^2 \mu^2}{4}. \label{aproxhvpoi}
\end{equation}
In the transformation of the above equation, we used a similar procedure to that used to derive Eq. (\ref{approxhhpoi}) with the relationship given by Eq. (\ref{gamma2}). 

\subsubsection{Single count rate}

If the idler detection information is not available to the signal side, the state of the signal photon can be expressed as a statistical mixture. 
When $x$ pairs are generated, the state of the signal is expressed as
\begin{equation}
\frac{1}{2^x} \bigotimes_{k=1}^x (|1,0\rangle_{sk} \langle1,0|_{sk} + |0,1\rangle_{sk} \langle0,1|_{sk}). 
\end{equation}
The state of the idler photon can also be expressed with a similar formula.
If we assume that the polarizer for the signal is set at {\it H} polarization, the probability of detecting the above state, $s(x)$, is given by
\begin{equation}
s(x) = \sum_{y=0}^x \frac{1}{2^x} \frac{x!}{y!(x-y)!} \{1-(1-\alpha)^{x-y}\}.
\end{equation}
Using $s(x)$, the single count rate $R_{sp}$ is expressed as
\begin{equation}
R_{sp} = \sum_{x=0}^\infty P_p (\mu,x) s(x)
\end{equation}
With the approximation $1-(1-\alpha)^x \simeq x \alpha$, the single count rate is reduced to
\begin{equation}
R_{sp} \simeq \frac{1}{2} \mu \alpha.
\end{equation}

\subsubsection{Visibility and contrast}

The visibility and the contrast with no approximation are obtained by plugging Eqs. (\ref{rhhpoi}) and (\ref{rhvpoi}) into the following equations.
\begin{eqnarray}
v_{p} &=& \frac{R_{HHp}-R_{HVp}}{R_{HHp}+R_{HVp}} \label{vpoifull} \\
\tau_p &=& \frac{R_{HHp}}{R_{HVp}}
\end{eqnarray}

With $1-(1-\alpha)^x \simeq x \alpha$, we can obtain the approximated visibility and contrast using Eqs. (\ref{approxhhpoi}) and (\ref{aproxhvpoi}) as 
\begin{eqnarray}
v_{p} &\simeq& \frac{1}{1+\mu} \label{vpoi}, \\
\tau_{p} &\simeq& 1+ \frac{2}{\mu} \label{tpoi}.
\end{eqnarray} 
Thus, we can ignore the loss dependence of $v_p$ and $\tau_p$ when $\alpha <<1$. 
It is obvious from the derivation shown above that the above approximated equations take account of both the limited number of multi-pair emission events and all the high order multi-pair emission events. 

\subsection{Indistinguishable entangled photon pairs}

Here we calculate the visibility of the TPI fringe obtained with indistinguishable entangled photon pairs, whose pair number distribution is given by Eq. (\ref{thermal}).

\subsubsection{Coincidence rate at fringe peak}

The detection probability when a single pair is created is the same as that for distinguishable pairs. 
However, the emergence probabilities of multi pairs are different from those for distinguishable pairs, since the number of patterns of polarization decreases because of the indistinguishability of the pairs. 
For example, let us consider the case when two pairs are created. 
There are only three patterns: both are {\it HH}, one is {\it HH} and the other is {\it VV}, both are {\it VV}. 
The patterns are summarized in Table 3. 

\begin{center}

Table 3: Two pairs \\

\begin{tabularx}{.60\linewidth}{ c|c| c }
\hline
Pattern & Emergence probability & Detection probability  
  \\
\hline
 Both {\it HH} & $\frac{1}{3}$ & $\{1-(1-\alpha)^2\}^2$   \\
\hline
 {\it HH} and {\it VV} & $\frac{1}{3}$ & $\alpha^2$   \\
\hline
Both {\it VV} & $\frac{1}{3}$ & 0   \\
\hline
\end{tabularx}
\end{center}

In general, when an $x$-pair state as shown by Eq. (\ref{indisxpair}) is created, the total number of patterns is given by $x+1$, and so the emergence probability of each pattern is given by $1/(x+1)$. 
If $y$ pairs are in the $VV$ state, the number of patterns of different sets of $HH$ and $VV$ pairs is simply 1, with a detection probability of $\{1-(1-\alpha)^{x-y}\}^2$. Thus, the coincidence probability in the setup when $x$ pairs are generated is given by
\begin{equation}
f_{indis}(x) = \sum_{y=0}^x \frac{1}{x+1} \{1-(1-\alpha)^{x-y}\}^2
\end{equation}

Since the probability of observing $x$ photon pairs is given by $P_{indis} (\mu,x)$ (Eq. (\ref{thermal})), the overall coincidence rate is obtained with the following equation.
\begin{equation}
R_{HHindis} = \sum_{x=0}^\infty f_{indis}(x) P_{indis} (\mu,x) \label{hhindis}
\end{equation}

When $1-(1-\alpha)^x \simeq x \alpha$, the above equation can be simplified as with distinguishable pairs. 
\begin{equation}
R_{HHindis} \simeq \sum_{x=0}^\infty \left\{ \sum_{y=0}^x \frac{1}{x+1} (x-y)^2 \alpha^2 P_{indis} (\mu,x) \right\} 
= \alpha^2 \left(\frac{\mu}{2}+\frac{\mu^2}{2} \right) \label{hhtapprox}
\end{equation}
In the transformation of the above equation, we used Eqs. (\ref{average}), (\ref{indis2}) and (\ref{appindistrans}).

\subsubsection{Coincidence rate at fringe bottom}

We now consider the coincidence rate at the bottom of the fringe, $R_{HV}$.  
With a similar calculation, the coincidence probability in the setup when $x$ pairs are generated is given by
\begin{equation}
g_{indis}(x) = \sum_{y=0}^x \frac{1}{x+1}  \{1-(1-\alpha)^{x-y}\}\{1-(1-\alpha)^{y}\}
\end{equation}

The overall coincidence rate at the fringe bottom is obtained with the following equation.
\begin{equation}
R_{HVindis} = \sum_{x=0}^\infty g_{indis}(x) P_{indis} (\mu,x)
\end{equation}

We can also derive a simpler expression for $R_{HVindis}$ when $1-(1-\alpha)^x \simeq x \alpha$. 
\begin{equation}
R_{HVindis} \simeq \sum_{x=0}^\infty \left\{\sum_{y=0}^x \frac{1}{x+1} (x-y) y \alpha^2 P_{indis} (\mu,x) \right\}
= \frac{\mu^2 \alpha^2}{4} \label{hvtapprox} 
\end{equation}
In the above equations, we used the relationship given by Eqs. (\ref{average}), (\ref{indis2}) and (\ref{appb}). 

\subsubsection{Single count rate}

When indistinguishable $x$ pairs are generated, the signal photon state can be expressed as the following statistical mixture.
\begin{equation}
\frac{1}{x+1} \sum_{k=0} |x-k,k\rangle_{sk} \langle x-k,k|_{xk}
\end{equation}
If the above state is detected with the polarizer set with H polarization, the detection probability $s_{indis} (x)$ is given by 
\begin{equation}
s_{indis} (x) = \frac{1}{1+x} \sum_{k=0}^x \{1-(1-\alpha)^k\}.
\end{equation}
The single count rate $R_{sindis}$ is given by
\begin{equation}
R_{sindis} = \sum_{x=0}^\infty P_{indis} (\mu,x) s_{indis} (x)
\end{equation}
With the approximation $1-(1-\alpha)^x \simeq x \alpha$, the above equation is simplified to
\begin{equation}
R_{sindis} \simeq \sum_{x=0}^\infty \left(\frac{P_{indis} (\mu,x)}{1+x} \sum_{k=0}^x k \alpha \right) = \frac{1}{2} \mu \alpha
\end{equation}
Thus, when the collection efficiency is small, the single count rates of the indistinguishable and distinguishable pairs are the same.

\subsubsection{Visibility and contrast}

The visibility and contrast for an indistinguishable source are given by

\begin{eqnarray}
v_{indis} &=& \frac{R_{HHindis}-R_{HVindis}}{R_{HHindis} + R_{HVindis}} \label{vindis} \\
\tau_{indis} &=& \frac{R_{HHindis}}{R_{HVindis}}
\end{eqnarray}

When $1-(1-\alpha)^x \simeq x \alpha$, the above values are simply obtained as
\begin{eqnarray}
v_{indis} &\simeq& \frac{\mu+2}{3 \mu +2} \label{vth} \\
\tau_{indis} &\simeq& 2+\frac{2}{\mu} \label{tth}
\end{eqnarray}

\begin{figure}[htb]
\centerline{\includegraphics[width=.6\linewidth]{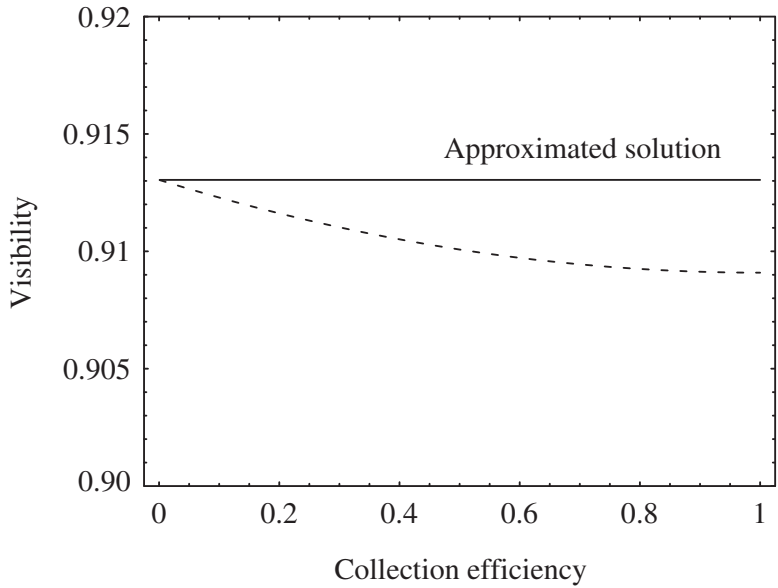}}

\caption{Visibility as a function of collection efficiency using Eqs. (\ref{vindis}) (dotted line) and (\ref{vth}) (solid line). } \label{123}
\end{figure}

Now we evaluate the validity of the visibility obtained with an approximation of $1-(1-\alpha)^x \simeq x \alpha$. Assuming $\mu=0.1$, Eqs. (\ref{vindis}) and (\ref{vth}) are plotted as a function of collection efficiency $\alpha$. The result is shown in Fig. \ref{123}. Here, we took account of up to 100 terms in the calculation of Eq. (\ref{vindis}). Although the deviation becomes larger as the collection efficiency increases, the difference in the visibility is only 0.4\% even with perfect collection efficiency. 
In experiments, it is very hard to obtain a collection efficiency exceeding 50\% even in the short wavelength band. Moreover, in most of the 1.5-$\mu$m band experiments, the collection efficiencies are typically a few per cent, largely because the single photon detectors for this band are inefficient.  
Also, we note that the difference between Eqs. (\ref{vindis}) and (\ref{vth}) becomes even smaller if we use a smaller $\mu$. Similar results are obtained with the formula for the distinguishable pairs. Thus, Eqs. (\ref{vpoi}) and (\ref{vth}) provide very good approximations of the visibilities.

\subsection{Comparison of distinguishable and indistinguishable pairs}

\begin{figure}[htb]

\centerline{\includegraphics[width=.6\linewidth]{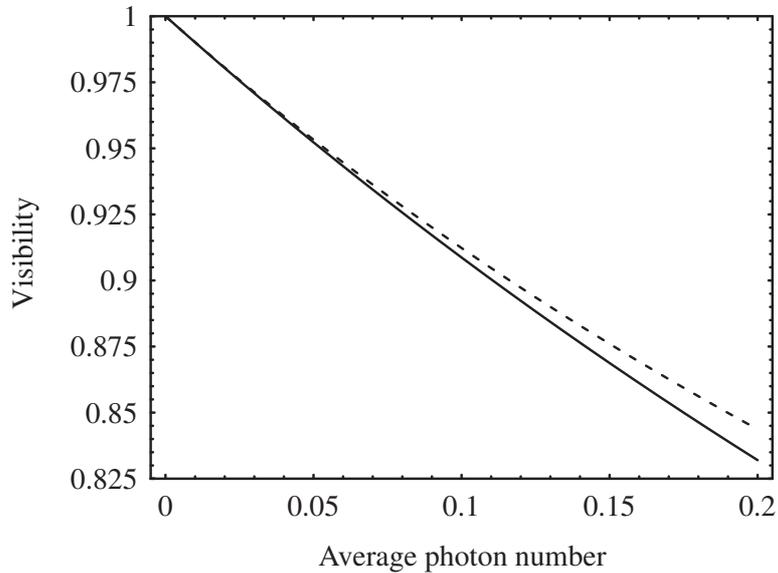}}

\caption{Visibility as a function of average photon number per pulse. Solid line: distinguishable pairs (Eq. (\ref{vpoifull})), dotted line: indistinguishable pairs (Eq. (\ref{vindis})). } \label{hikaku}
\end{figure}

We compared the visibilities of distinguishable and indistinguishable entangled photon pairs using Eqs. (\ref{vpoifull}) and (\ref{vindis}), where we assumed $\alpha=0.1$ and took account of up to 100 terms in the calculations of these equations. 
The result is shown in Fig. \ref{hikaku}. This indicates that the indistinguishable pairs exhibit slightly better visibility than the distinguishable pairs. For example, at $\mu=0.1$, the visibilities for the indistinguishable and distinguishable pairs are 91.2\% and 90.8\%, respectively. We also note that we can obtain almost the same results if we use approximated equations (\ref{vpoi}) and (\ref{vth}). 
The result reflects the fact that the contrast of indistinguishable pairs $\tau_{indis}$ is greater than that of distinguishable pairs $\tau_p$ by $\sim 1$, as we saw in the approximated equations of the contrasts given by Eqs. (\ref{tpoi}) and (\ref{tth}). 
In addition, Eqs. (\ref{approxhhpoi}) and (\ref{hhtapprox}) suggest that the indistinguishable pairs yield a larger peak coincidence rate than the distinguishable pairs. These results mean that, in an entanglement-based QKD system, an indistinguishable entangled pair source gives a larger key rate with a smaller error rate than a distinguishable source, which yields a higher secure key rate. Thus, the presented analysis clearly confirms that an indistinguishable entangled photon pair source performs better than a distinguishable one in quantum information experiments. 

\section{Density matrix reconstruction using quantum state tomography} \label{qst}

The method for calculating the coincidence rate described in the previous section can be used to calculate the 16 different combinations of two-photon projection measurements shown in Table 4. In this table, $|H\rangle$ ($|V\rangle$) denotes a state in which there is a photon in the H (V) polarization mode. Other states are defined as $|+\rangle = (|H\rangle + |V\rangle)/\sqrt{2}$, $|L\rangle = (|H\rangle + i|V\rangle)/\sqrt{2}$, and $|R\rangle = (|H\rangle -i |V\rangle)/\sqrt{2}$. By using these coincidence rates, we can construct the density matrix of the two-photon state using quantum state tomography (QST) as reported in \cite{dfv}.  Here, again we assume that only the accidental coincidences caused by multi-photon emission events are the origin of the degradation in the entanglement quality, and there is no other cause for degradation. 

\newpage
\begin{center}
Table 4. Assumed projected state. \\
\vspace{3mm}
\begin{tabularx}{.30\linewidth}{ c|c| c }
\hline
$\nu$ & Photon 1 & Photon 2 
  \\
\hline
1 & $|H\rangle$ & $|H\rangle$   \\
\hline
2 & $|H\rangle$ & $|V\rangle$   \\
\hline
3 & $|V\rangle$ & $|V\rangle$  \\
\hline
4 & $|V\rangle$ & $|H\rangle$  \\
\hline
5 & $|R\rangle$ & $|H\rangle$    \\
\hline
6 & $|R\rangle$ & $|V\rangle$  \\
\hline
7 & $|+\rangle$ & $|V\rangle$   \\
\hline
8 & $|+\rangle$ & $|H\rangle$   \\
\hline
9 & $|+\rangle$ & $|R\rangle$   \\
\hline
10 & $|+\rangle$ & $|+\rangle$   \\
\hline
11 & $|R\rangle$ & $|+\rangle$    \\
\hline
12 & $|H\rangle$ & $|+\rangle$  \\
\hline
13 & $|V\rangle$ & $|+\rangle$    \\
\hline
14 & $|V\rangle$ & $|L\rangle$   \\
\hline
15 & $|H\rangle$ & $|L\rangle$   \\
\hline
16 & $|R\rangle$ & $|L\rangle$    \\
\hline
\end{tabularx}
\end{center}


$\nu=1$, 3, 10 and 16 correspond to positively correlated events. Therefore, it is easy to show that these events have the same coincidence rate as $R_{HH}$ calculated in the previous section. Similarly, the coincidence rates for $\nu=$2, 4, which correspond to negatively correlated events, are given by $R_{HV}$. From the symmetry of the equations, the coincidence rates for other basis mismatch cases have the same value. Therefore, if we can calculate a coincidence rate for one of those combinations, we can obtain the rates for all $\nu$ values and thus construct the density matrix. In the following, we consider the $\nu=12$ case, in which the signal photon is projected to the $|H\rangle$ state and the idler photon is projected to the $|+\rangle$ state.

\subsection{Distinguishable pairs}

Assume that the projection measurement for the signal photon is with {\it HV} basis, and that for the idler is with diagonal basis. The single pair state is rewritten as
\begin{eqnarray}
|\Phi^+\rangle &=& \frac{1}{\sqrt{2}} (|H\rangle_s |H\rangle_i + |V\rangle_s |V\rangle_i ) \nonumber \\
&=& \frac{1}{2} (|H\rangle_s |+\rangle_i + |V\rangle_s |+\rangle_i + |H\rangle_s |-\rangle_i - |V\rangle_s |-\rangle_i).
\end{eqnarray}

When an $x$-pair state given by Eq. (\ref{pxpair}) is generated, 
the coincidence probability is given by
\begin{equation}
h(x) = \left(\sum_{k=0}^x \frac{1}{2^x} \frac{x!}{k! (x-k)!} \{1-(1-\alpha)^k\} \right) \left(\sum_{j=0}^x \frac{1}{2^x} \frac{x!}{j! (x-j)!} \{1-(1-\alpha)^j\} \right)\end{equation}
The overall probability of coincidence is obtained as
\begin{equation}
R_{H+p} = \sum_{x=0}^\infty h(x) P_{px} (\mu). \label{h+pfull}
\end{equation}
$r_\nu$ is then obtained using coincidence rates of {\it HH}, {\it HV} and {\it H}+ projection measurements.
\begin{equation}
r_\nu = \left\{ 
\begin{array}{rl}
R_{HHp} & \nu=1,3,10,16 \\
R_{HVp} & \nu=2,4 \\
R_{H+p} & \nu=\mbox{others}
\end{array}
\right.
\end{equation}

When $1-(1-\alpha)^k \simeq k \alpha$, Eq. (\ref{h+pfull}) can be simplified to \begin{equation}
R_{H+p} \simeq \alpha^2 \left(\frac{\mu}{4}+\frac{\mu^2}{4}\right)
\end{equation}
In the following, we calculate the density matrix assuming $1-(1-\alpha)^k \simeq k \alpha$. Let us denote the coincidence rate for each $\nu$ value in Table 4 as $r_\nu$. Using the results of the above equation as well as those of the previous section, $r_\nu$ is obtained as follows. 
\begin{equation}
r_\nu \simeq \left\{ 
\begin{array}{rl}
\alpha^2 \left(\frac{\mu}{2} + \frac{\mu^2}{4} \right) & \nu=1,3,10,16 \\
\frac{\alpha^2\mu^2}{4} & \nu=2,4 \\
\alpha^2 \left(\frac{\mu}{4} + \frac{\mu^2}{4} \right) & \nu=\mbox{others}
\end{array}
\right.
\end{equation}

\begin{figure}[htb]

\centerline{\includegraphics[width=.9\linewidth]{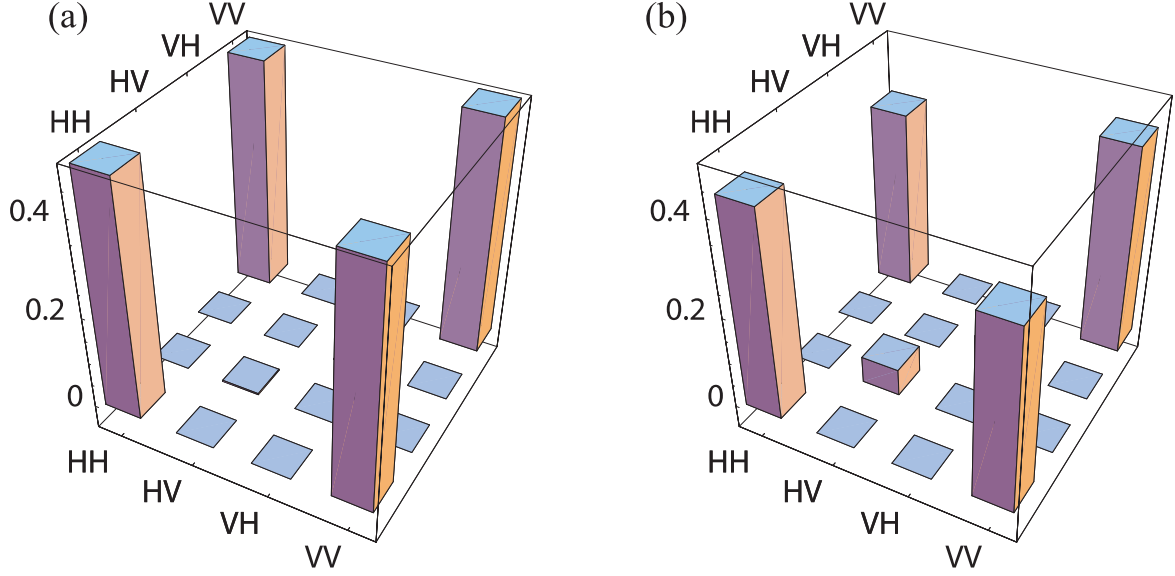}}

\caption{Graphical representation of density matrix of distinguishable photon-pair states. (a) $\mu=0.01$, (b) $\mu=0.3$. } \label{dm}
\end{figure}

Using $r_\nu$ shown above and the procedure presented in \cite{dfv}, we can obtain the following density matrix.

\begin{equation}
\rho_p (\mu) \simeq 
\left(
\begin{array}{cccc}
\frac{2 +\mu}{4 + 4 \mu} & 0 & 0 & \frac{1}{2+2\mu} \cr
0 & \frac{\mu}{4 + 4 \mu} & 0 & 0 \cr
0 & 0 & \frac{\mu}{4 + 4 \mu} & 0 \cr
\frac{1}{2+2\mu} & 0 & 0 & \frac{2 +\mu}{4 + 4 \mu}
\end{array}
\right)
\end{equation}
Thus, the density matrix is obtained as a function of $\mu$. 
The density matrix is graphically shown in Fig. \ref{dm} for $\mu=0.01$ and 0.3. 
Using this density matrix, the concurrence can also be obtained as a simple function of $\mu$, by using the procedure described in \cite{dfv}. The concurrence is 
\begin{equation}
Conc_p \simeq \frac{2-\mu}{2(\mu+1)}. \label{cuncp}
\end{equation}

\subsection{Indistinguishable pairs}

Here we undertake a projection measurement on the indistinguishable pairs in the way described in the previous subsection.

Let us assume that we have an $x$ indistinguishable pair state, whose state is given by Eq. (\ref{indisxpair}). 
In that joint state, $|x-k,k\rangle_i$ can be rewritten as follows. 
\begin{eqnarray}
|x-k,k\rangle_i &=& \frac{1}{\sqrt{(x-k)!k!}} (a_{iH}^\dagger)^{x-k} (a_{iV}^\dagger)^k |0\rangle_i \nonumber \\
&=& \frac{1}{(\sqrt{2})^x \sqrt{(x-k)!k!}} (a_{i+}^\dagger + a_{i-}^\dagger)^{x-k} (a_{i+}^\dagger - a_{i-}^\dagger)^{k} |0\rangle_i \label{x-k}
\end{eqnarray}
Here, we used the unitary transformations between HV and the diagonal basis given by
\begin{eqnarray}
a_H^\dagger &=& \frac{1}{\sqrt{2}} (a_+^\dagger + a_-^\dagger), \\
a_V^\dagger &=& \frac{1}{\sqrt{2}} (a_+^\dagger - a_-^\dagger). 
\end{eqnarray}
Using binominal theorem $(a+b)^k = \sum_{n=0}^k ._k C_n a^n b^{k-n}$, Eq. (\ref{x-k}) is further transformed as
\begin{equation}
|x-k,k\rangle_i = \frac{1}{(\sqrt{2})^x \sqrt{(x-k)!k!}} \left\{ \sum_{m=0}^{x-k} ._{x-k} C_m (a_{i+}^\dagger)^m (a_{i-}^\dagger)^{x-k-m} \right\} \left\{ \sum_{n=0}^{k} ._{k} C_n (a_{i+}^\dagger)^n (-a_{i-}^\dagger)^{k-n} \right\} |0\rangle_i \end{equation}
From the above equation, the coincidence probability of the $x$ indistinguishable pair state is obtained as
\begin{eqnarray}
& &h_{indis} (\mu, \alpha, x) \nonumber \\ 
&=& \frac{1}{x+1} \sum_{k=0}^{x-1} \left[
\frac{1}{2^x (x-k)! k!} \sum_{m=0}^{x-k} \sum_{n=0}^k \frac{\{(x-k)!\}^2}{m!(x-k-n)!} \frac{(k!)^2}{n!(k-n)!}
\{1-(1-\alpha)^{n+m}\} \{1-(1-\alpha)^{x-k}\}
 \right]. 
\end{eqnarray}
Using $h_{indis}$, we can obtain the overall coincidence rate for the $|H\rangle_s$ and $|+\rangle_i$ bases with the following equation.
\begin{equation}
R_{H+indis} (\mu,\alpha) = \sum_{x=0}^\infty P_{indis}(\mu,x) h_{indis} (\mu, \alpha, x) \label{h+}
\end{equation}

$r_\nu$ is then determined using coincidence rates obtained with $HH$, $HV$ and $H+$ projection measurements.
\begin{equation}
r_\nu = \left\{ 
\begin{array}{rl}
R_{HHindis} & \nu=1,3,10,16 \\
R_{HVindis} & \nu=2,4 \\
R_{H+indis} & \nu=\mbox{others}
\end{array}
\right.
\end{equation}
The density matrix, and consequently the concurrence, are obtained using those rates as a function of $\mu$ and $\alpha$.

When $1-(1-\alpha)^k \simeq k \alpha$, Equation (\ref{h+}) is approximated as follows. 
\begin{equation}
R_{H+indis} (\mu, \alpha) \simeq \alpha^2 \left(\frac{3\mu^2}{8} + \frac{\mu}{4}\right) \label{h+tapprox}
\end{equation}

By using the approximated coincidence rate equations, $r_\nu$ is now obtained as
\begin{equation}
r_\nu \simeq \left\{ 
\begin{array}{rl}
\alpha^2 \left(\frac{\mu^2}{2}+\frac{\mu}{2} \right) & \nu=1,3,10,16 \\
\frac{\alpha^2 \mu^2}{4} & \nu=2,4 \\
\alpha^2 \left(\frac{3 \mu^2}{8} + \frac{\mu}{4}\right)  & \nu=\mbox{others}
\end{array}
\right.
\end{equation}
With those $r_\nu$ values, the density matrix is obtained as
\begin{equation}
\rho_{indis} (\mu) \simeq 
\left(
\begin{array}{cccc}
\frac{1 +\mu}{2 + 3 \mu} & 0 & 0 & \frac{2+\mu}{4+6\mu} \cr
0 & \frac{\mu}{4 + 6 \mu} & 0 & 0 \cr
0 & 0 & \frac{\mu}{4 + 6 \mu} & 0 \cr
\frac{2 + \mu}{4+6\mu} & 0 & 0 & \frac{1 +\mu}{2 + 3 \mu}
\end{array}
\right)
\end{equation}
Using the above density matrix, we approximate the concurrence with the following equation. 
\begin{equation}
Conc_t (\mu) \simeq \frac{2}{2+3 \mu} \label{cunct}
\end{equation}

\begin{figure}[htb]

\centerline{\includegraphics[width=.6\linewidth]{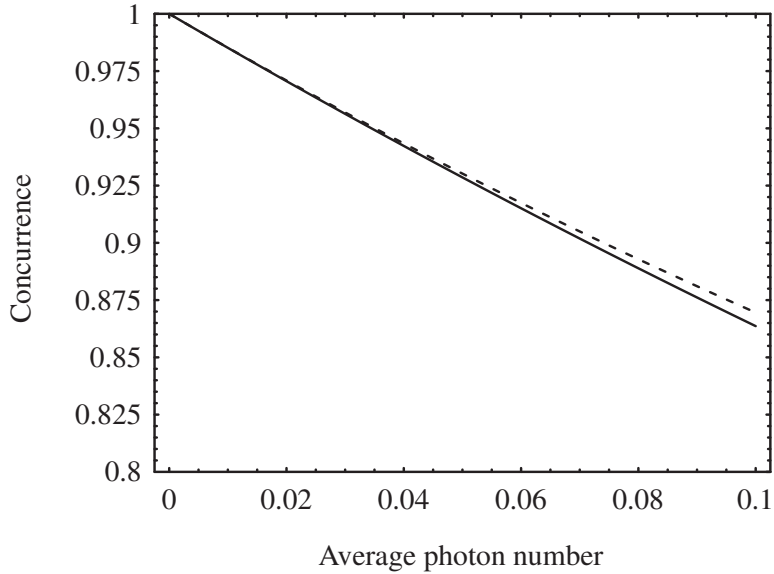}}

\caption{Concurrence as a function of average photon-pair number per pulse $\mu$. Solid line, distinguishable pairs (Eq. (\ref{cuncp})), dotted line, indistinguishable pairs (Eq. (\ref{cunct})). } \label{cunc}
\end{figure}

Using Eqs. (\ref{cuncp}) and (\ref{cunct}), we plot the concurrences of distinguishable and indistinguishable entangled photon pairs as a function of $\mu$ in Fig. \ref{cunc}. Thus, indistinguishable photon pairs have a slightly better concurrence than distinguishable ones, as expected from the visibility calculations provided in the previous section. 


\section{Correction terms by dark count} \label{dc}

From the results described in the previous sections, it is clear that visibility and concurrence can be made infinitely close to unity simply by decreasing $\mu$. 
However, the dark counts of detectors limit those values in real experiments. 
Here we include the dark count in the theory developed in the previous sections. 

We assume that $x$ photons are incident on a threshold detector with a collection efficiency $\alpha$ and a dark count probability $d$. When $\alpha$ and $d$ are both much smaller than 1, the detection probability is approximately given by $x \alpha + d$. 
With this approximation, we can calculate the coincidence rates in the previous sections by replacing $x \alpha$ with $x \alpha + d$. 

In general, the two detectors have different collection efficiencies and dark count probabilities. We now denote their signal and idler values with subscripts $s$ and $i$. 

For example, the coincidence rate of an {\it HH} measurement on indistinguishable pairs is given by 
\begin{eqnarray}
R_{HHindis} &\simeq& \sum_{x=0}^\infty \left\{\sum_{y=0}^x \frac{P_{thx (\mu)}}{x+1} \{(x-y)\alpha_s + d_s\} \{(x-y) \alpha_i + d_i\} \right\} \nonumber \\
&=& \sum_{x=0}^\infty \left\{\sum_{y=0}^x \frac{P_{thx (\mu)}}{x+1} \{(x-y)^2 \alpha_s \alpha_i + (x-y) \alpha_s d_i + (x-y) \alpha_i d_s + d_s d_i \} \right\} \nonumber \\
&=& \alpha_s \alpha_i \left(\frac{\mu^2}{2} + \frac{\mu}{2}\right) + \frac{\mu \alpha_s d_i}{2} + \frac{\mu \alpha_i d_s}{2} + d_s d_i
\end{eqnarray}
The first term corresponds to the result shown by Eq. (\ref{hhtapprox}). The second and the third terms give the coincidence between counts caused by a photon and a dark count, and the fourth is the coincidence between dark counts. With a similar procedure, we can include the dark counts for all the above equations. In Appendix \ref{appdc}, we summarize the equations that take account of the dark counts. 
We can optimize $\mu$ to obtain the maximum visibility and concurrence using these equations.

\section{Time-bin entangled photon pairs} \label{tb}

In this section, we derive the equations for time-bin entangled photon pairs \cite{brendel}. 
A time-bin entangled state is expressed with the following equation. 
\begin{equation}
|\Psi \rangle = \frac{1}{\sqrt{2}}(|1\rangle_s |1\rangle_i + |2\rangle_s |2\rangle_i) \label{tent}
\end{equation}
Here, $|k\rangle_z$ denotes a state in which there is a photon in the $k$th time slot and in a mode $z$. 
We can generate a time-bin entangled photon pair by pumping a nonlinear optical medium with a coherent double pulse \cite{brendel,takesue3,longd}. Therefore, the interaction Hamiltonian of the spontaneous parametric process that generates time-bin entanglement is exactly same as Eq. (\ref{inthv}), if we replace the {\it H} and {\it V} polarization modes with the first and second pulses, respectively. As a result, the pair number distribution of the indistinguishable time-bin entangled photon pairs is the same as that of the polarization entangled photon pairs given by Eq. (\ref{thermal}). Similarly, the number distribution of the distinguishable pairs is Poissonian as given by Eq. (\ref{poi}). 

In the measurement, each photon is passed through a 1-bit delayed Mach-Zehnder interferometer. If we denote the two output ports of the interferometer as $a$ and $b$, the state $|k\rangle_z$ can be transformed by the interferometer as follows. 
\begin{equation}
|k\rangle_z \to \frac{1}{2} \left(|k,a\rangle_z - |k,b\rangle_z + e^{i\theta_z} |k+1,a\rangle_z + e^{i\theta_z} |k+1,b \rangle_z \right) \label{henkan}
\end{equation}
Here, $|k,x\rangle_z$ indicates a state where there is a photon in the $k$th time slot, the output $x$, and the mode $z$. With this equation, Eq. (\ref{tent}) evolves into the following state by passing through the interferometers. 
\begin{eqnarray}
|\Psi \rangle &\to& \frac{1}{4 \sqrt{2}} \left\{|1,a\rangle_s |1, a\rangle_i - |1,a\rangle_s |1,b\rangle_i - |1,b\rangle_s |1, a\rangle_i + |1,b\rangle_s |1,b\rangle_i \right.  \nonumber \\
& & + \{e^{i(\theta_s+\theta_i)}+1\}|2,a\rangle_s |2, a\rangle_i +  \{e^{i(\theta_s+\theta_i)}-1\}|2,a\rangle_s |2,b\rangle_i\nonumber \\
& & + \{e^{i(\theta_s+\theta_i)}-1\}|2,b\rangle_s |2, a\rangle_i +  \{e^{i(\theta_s+\theta_i)}+1\}|2,b\rangle_s |2,b\rangle_i \nonumber \\
& & \left. +e^{i(\theta_s+\theta_i) }|3,a\rangle_s |3, a\rangle_i + e^{i(\theta_s+\theta_i)}|3,a\rangle_s |3,b\rangle_i + e^{i(\theta_s+\theta_i)}|3,b\rangle_s |3,a\rangle_i + e^{i(\theta_s+\theta_i)}|3,b\rangle_s |3,b\rangle_i
\right\}, \label{1bit}
\end{eqnarray}
Here, the terms that do not contribute to coincidences are not shown. When only ports $a$ are connected to photon detectors, a constructive TPI occurs when $\theta_s + \theta_i = 2 n \pi$ ($n$: integer). With a 100\% collection efficiency, the coincidence rate per pair is 1/8. 
Therefore, with the same average pair number, the coincidence probability of the time-bin entangled photon pairs is 1/4 of that of polarization entangled pairs. Here, note that the average pair number for the time-bin entangled photon pairs is defined as the number per two temporal modes. 
The decrease of the coincidence probability comes from the fact that the loss for each photon effectively increases by 3 dB because the photon is observed in the 2nd time slot with a 50\% probability. Therefore, by replacing $\alpha$ with $\alpha/2$ in the equations in section 3, we can obtain the formulas for time-bin entanglement.

Let us describe an example. 
In a TPI measurement of a time-bin entanglement, the peak of the fringe is obtained when the detectors are both set at port $a$ or $b$. So we denote the peak coincidence rate as $R_{aa}$. Similarly, the coincidence rate at the bottom of the fringe is denoted as $R_{ab}$.  
With the presence of dark counts, the TPI peak coincidence rate of the indistinguishable time-bin entanglement $R_{aaindis}$ is calculated as
\begin{equation}
R_{aaindis} = \sum_{x=0}^\infty \left(\sum_{y=0}^x \frac{P_{indis} (\mu,x)}{x+1} \left\{1-\left(1-\frac{\alpha}{2}\right)^{x-y} \right\}^2  \right),  
\end{equation}
which is 1/4 of $R_{HHindis}$ given by Eq. (\ref{hhindis}). With an approximation $1-(1-\alpha/2)^x \simeq x \alpha/2$ and with non-negligible dark counts, $R_{peakindis}$ is obtained as
\begin{eqnarray}
R_{aaindis}&\simeq& \sum_{x=0}^\infty \left\{\sum_{y=0}^x \frac{P_{thx (\mu)}}{4(x+1)} \left\{(x-y)\frac{\alpha_s}{2} + d_s\right\} \left\{(x-y) \frac{\alpha_i}{2} + d_i\right\} \right\} \nonumber \\
&=& \alpha_s \alpha_i \left(\frac{\mu^2}{8} + \frac{\mu}{8}\right) + \frac{\mu \alpha_s d_i}{4} + \frac{\mu \alpha_i d_s}{4} + d_s d_i.
\end{eqnarray}
The coincidence rates for other cases are summarized in Appendix \ref{apptb}.

\begin{figure}[htb]

\centerline{\includegraphics[width=.9\linewidth]{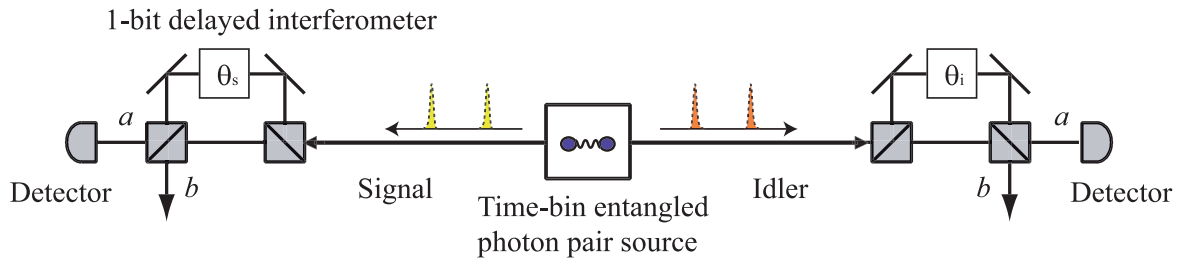}}

\caption{Setup for time-bin entangled photon-pair measurement. } \label{tbfig}
\end{figure}


\section{Summary} \label{summary}

We described a theoretical analysis of the quality of entangled photon pairs generated by a spontaneous parametric process. Assuming the use of threshold detectors, we derived simple formulas for the TPI visibilities for both distinguishable and indistinguishable polarization entangled photon pairs. 
The main results obtained in this paper are summarized below. 
\begin{enumerate}
\item We presented the exact formulas for the coincidence rates of the peak and bottom of TPI fringes, taking account of the accidental coincidences caused by all the high-order multi-pair emissions. With these formulas, we can calculate the visibilities as a function of the average pair number and the photon collection efficiencies. In addition, these somewhat complex formulas can be reduced to simple functions of the average pair number when the collection efficiencies are small. These approximated formulas also include the effects of all the high-order multi-pair emissions. 
\item Using the derived equations, we compared the visibilities of the distinguishable and indistinguishable entangled pairs. The result suggests that the indistinguishable pairs provide improved visibility with the same average pair number, implying that we can construct quantum information systems with better performance using the indistinguishable entangled photon-pair sources. 
\item The full density matrix is obtained as a function of the average pair number and the collection efficiencies. Here, too, simpler formulas for the density matrix were presented. In addition, we derived simple formulas of concurrences as a function of the average pair number for distinguishable and indistinguishable pairs. 
\item We included the effect of dark counts in the theory, which enabled us to predict the visibility experimentally obtained with a specific average photon number. 
\item We modified the above formulas for time-bin entangled photon pairs. 
\end{enumerate}

 We believe that the contents of this paper will be useful for designing and evaluating photonic quantum information experiments with entangled photon-pair sources based on spontaneous parametric processes. 

\section{Acknowledgement}

H. T. acknowledges that he was strongly motivated to undertake this research by discussion with H. C. Lim of The University of Tokyo.

\appendix

\section*{Appendices}

\section{Coincidence-to-accidental measurement of photon-pair state} \label{car}

\begin{figure}[htb]

\centerline{\includegraphics[width=.9\linewidth]{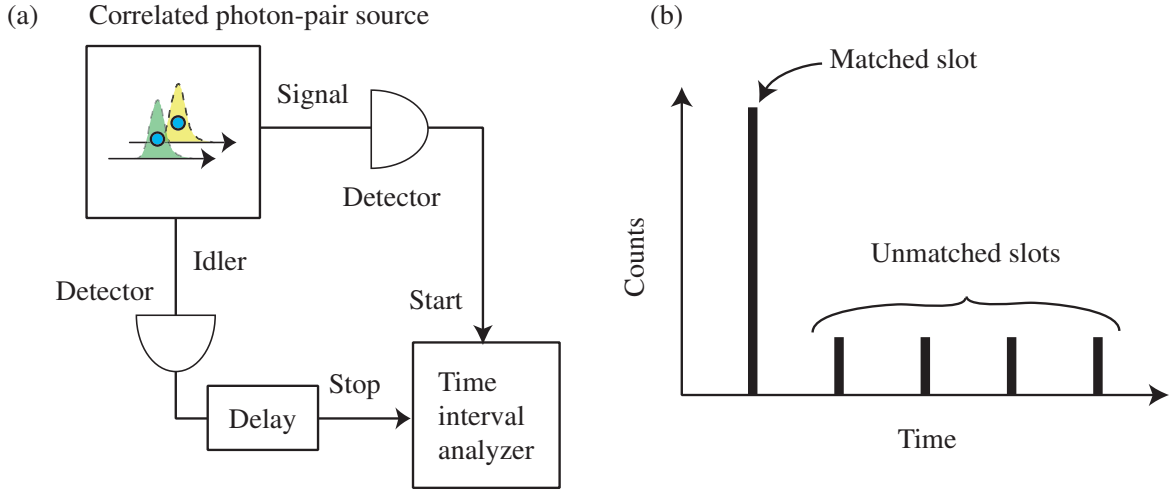}}

\caption{(a) CAR measurement setup, (b) typical histogram of coincidence events. } \label{carsetup}
\end{figure}


This appendix describes the coincidence-to-accidental ratio (CAR) measurements, which is used to evaluate correlated photon pairs (i.e. a source that generates a product state but not an entangled state) \cite{fio,takesue2}. We derive the theoretical CAR for distinguishable and indistinguishable correlated pair sources. 

A typical CAR measurement setup is shown in Fig. \ref{carsetup} (a). 
A correlated photon pair source outputs signal-idler photon pairs, which exhibit a temporal correlation. Each photon is detected by a threshold detector. The detection signal from the detector for the signal photon is used as a start pulse for the TIA. The idler detection signal is delayed by an electrical delay circuit and then used as a stop pulse. We insert the delay so that the coincidence events caused by the photons generated at the same time do not fall within the TIA deadtime. Then, we can obtain a histogram of coincidence events as a function of time, as shown in Fig. \ref{carsetup} (b). If there is a temporal correlation between a signal and an idler photon, we observe a peak in the histogram in the temporal position that corresponds to the coincidence events caused by the photons generated at the same temporal position. We call the time slot where this peak is located a ``matched slot". We also observe other side peaks, which are the coincidence events induced by the photons generated at different temporal positions. We call these time slots ``unmatched slots". 
 
We assume that the probability of obtaining $x$ pairs is given by $P (\mu,x)$. 
Then, the count rate per pulse at each detector is given by
\begin{equation}
Ct(\mu) = \sum_{x}^\infty P(\mu,x) \{1-(1-\alpha)^x\}. 
\end{equation}
The coincidence rate at an unmatched slot $R_{um}$ is simply given by the coincidence probability of two uncorrelated events, which is expressed as
\begin{equation}
R_{um} = \left( \sum_{x}^\infty P (\mu,x) \{1-(1-\alpha)^x\} \right)^2. \label{um}
\end{equation}
On the other hand, the coincidence rate at the matched slot $R_{m}$ is given by
\begin{equation}
R_{m} = \sum_x^\infty P (\mu,x) \{1-(1-\alpha)^x\}^2. \label{m}
\end{equation}
The CAR $C$ is given by
\begin{equation}
C = \frac{R_{m}}{R_{um}}. \label{ccc}
\end{equation}

\subsection{Distinguishable pairs}

When the source generates distinguishable correlated pairs, it is easy to show that the pair number distribution is Poissonian and given by Eq. (\ref{poi}) as in the case of entangled pairs, using a similar procedure to that described in the previous section. 
We can obtain the CAR for distinguishable pairs simply by replacing $P (\mu,x)$ with $P_{p} (\mu,x)$ in Eqs. (\ref{um}), (\ref{m}) and (\ref{ccc}).  

When the collection efficiency is small, $C$ can be approximated in a much simpler form. 
If $\alpha << 1$, the detection probability of $x$ photons $\{1-(1-\alpha)^x\}$ can be approximated with $x \alpha$. Then, using Eq. (\ref{average}), $R_{um}$ can be simplified to
\begin{equation}
R_{um} \simeq  \left( \sum_{x}^\infty P_p (\mu,x) x \alpha \} \right)^2 = \mu^2 \alpha^2.
\end{equation}
Also, using Eq. (\ref{p2}), $R_{um}$ is transformed to 
\begin{equation}
R_{m} \simeq \sum_{x}^\infty P_p (\mu,x) x^2 \alpha^2. 
=\alpha^2 \mu (\mu+1)
\end{equation}
Thus, CAR for distinguishable correlated pairs $C_p$ is given by
\begin{equation}
C_{p} = 1+\frac{1}{\mu}. \label{cpoi}
\end{equation}
This result agrees with the result obtained with the simple Poissonian model introduced in \cite{takesue2}.

\subsection{Indistinguishable pairs}

When the source generates indistinguishable correlated pairs, the pair number distribution is slightly different from that for the indistinguishable entangled pairs given by Eq. (\ref{thermal}). Using the interaction Hamiltonian given by Eq. (\ref{intpp}) and a similar procedure to that described in section 2.1, the probability of generating $x$ pairs $P_{th} (x)$ is given by
\begin{equation}
P_{th} (x) = \frac{\Gamma^{2x}}{C^2} = \frac{\tanh^{2x} \chi t}{\cosh^2 \chi t}
\end{equation}
It is straightforward to show that $P_{th} (x)$ can be rewritten with the average pair number $\mu$ as
\begin{equation}
P_{th} (\mu,x) = \frac{\mu^x}{(1+\mu)^{x+1}}. \label{th}
\end{equation}
Thus, the number distribution of indistinguishable correlated photon pairs is expressed as the thermal distribution. The CAR for the indistinguishable pairs is obtained by plugging $P_{th} (\mu,x)$ into Eq. (\ref{ccc}). 

Let us consider the case when $\alpha << 1$. 
$R_{um}$ can be approximated as
\begin{equation}
R_{side} \simeq \left(\sum_{x=0}^\infty P_{th} (\mu,x) \alpha x\right)^2 = \mu^2 \alpha^2,
\end{equation}
where we used Eq. (\ref{average}). 
Similarly, using Eq. (\ref{th2}) found in Appendix \ref{appcal}, we can calculate $R_{m}$ as
\begin{equation}
R_{m} \simeq \sum_{x=0}^\infty P_{th} (\mu,x) x^2 \alpha^2 = \alpha^2 (2 \mu^2 + \mu)
\end{equation}
Thus, CAR is obtained with the following equation.
\begin{equation}
C_{th} = 2 + \frac{1}{\mu} \label{cth}
\end{equation}
With Eqs. (\ref{cpoi}) and (\ref{cth}), the CAR obtained with a indistinguishable correlated photon-pair source is larger by 1 than that obtained with a distinguishable source.

\section{Details of calculation in section \ref{tpi}} \label{appcal}

The pair number distributions $P_{indis} (\mu,x)$, $P_{p} (\mu,x)$ and $P_{th} (\mu,x)$, which are defined by Eqs. (\ref{thermal}), (\ref{poi}) and (\ref{th}), respectively, satisfy the following equations. 
\begin{eqnarray}
& & \sum_{x=0}^\infty P_p (\mu,x) = \sum_{x=0}^\infty  P_{indis} (\mu,x) = \sum_{x=0}^\infty  P_{th} (\mu,x) = 1 \\
& & \sum_{x=0}^\infty x P_p (\mu,x) = \sum_{x=0}^\infty  x P_{indis} (\mu,x) = \sum_{x=0}^\infty  x P_{th} (\mu,x) = \mu \label{average} \\
& & \sum_{x=0}^\infty x^2 P_{p} (\mu,x) = \mu +  \mu^2 \label{p2} \\
& & \sum_{x=0}^\infty x^2 P_{indis} (\mu,x) = \mu + \frac{3}{2} \mu^2 \label{indis2} \\
& & \sum_{x=0}^\infty x^2 P_{th} (\mu,x) = \mu + 2 \mu^2 \label{th2}
\end{eqnarray}

The derivation of Eq. (\ref{approxhhpoi}) is given in the following. 
\begin{eqnarray}
R_{HHp} &\simeq& \alpha^2 \sum_{x=0}^\infty \left\{
\sum_{y=0}^x \frac{1}{2^x} \frac{x!}{y! (x-y)!} (x-y)^2 \frac{\mu^x}{x!} e^{-\mu}  \right\} \nonumber \\
&=& \frac{\alpha^2 \mu e^{-\mu}}{2} + \alpha^2 \sum_{x=2}^\infty \left\{
\frac{\mu^x}{2^x} e^{-\mu} \sum_{y=0}^x \frac{x-y}{y!(x-y-1)!} \label{temp}
\right\}
\end{eqnarray}
When $x > 1$, we can use the following relationship. 
\begin{equation}
\sum_{y=0}^x \frac{x-y}{y!(x-y-1)!} = \frac{2^{x-1} x}{(x-1)!} - \frac{2^{x-2}}{(x-2)!},
\end{equation}
Using the above, Eq. (\ref{temp}) can be transformed as
\begin{eqnarray}
R_{HHp} &\simeq& \frac{\alpha^2 \mu e^{-\mu}}{2}+ \alpha^2 \sum_{x=2}^\infty \left\{
\frac{\mu^x}{2^x} e^{-\mu} \left(\frac{2^{x-1} x}{(x-1)!} - \frac{2^{x-2}}{(x-2)!} \right) 
\right\} \nonumber \\
&=&  \frac{\alpha^2 \mu e^{-\mu}}{2}+ \alpha^2 \sum_{x=2}^\infty \left\{
\frac{1}{2} \frac{\mu^x}{(x-1)!}e^{-\mu} x - \frac{1}{4} \frac{\mu^x}{(x-2)!} e^{-\mu} 
\right\} \nonumber \\
&=& \frac{\alpha^2 \mu e^{-\mu}}{2}+ \frac{\alpha^2 \mu}{2} \sum_{x=0}^\infty \left\{e^{-\mu} \frac{\mu^{x-1}}{(x-1)!} (x-1) + e^{-\mu} \frac{\mu^{x-1}}{(x-1)!} \right\} - \frac{\alpha^2 \mu^2}{4} \sum_{y=0}^x \left(\frac{\mu^{x-2}}{(x-2)!} e^{-\mu} \right) \nonumber \\
&=& \frac{\alpha^2 \mu}{2} \sum_{x=2}^\infty \left\{e^{-\mu} \frac{\mu^{x-1}}{(x-1)!} (x-1)\right\} +  \frac{\alpha^2 \mu}{2} \left\{e^{-\mu}+ \sum_{x=2}^\infty e^{-\mu} \frac{\mu^{x-1}}{(x-1)!} \right\} - \frac{\alpha^2 \mu^2}{4} \sum_{y=0}^x \left(\frac{\mu^{x-2}}{(x-2)!} e^{-\mu} \right) \nonumber \\
&=& \alpha^2 \left( \frac{\mu}{2} + \frac{\mu^2}{4}\right) \label{apphhp}
\end{eqnarray}

Also, the following relationships are used in the calculations in section \ref{tpi}. The first equation assumes that $x$ is an integer that is larger than 1. 

\begin{equation}
\sum_{y=0}^x \frac{1}{(y-1)!(x-y-1)!} = \frac{2^{x-2}}{(x-2)!} \label{gamma2}
\end{equation}
\begin{equation}
\sum_{y=0}^x \frac{1}{x+1} (x-y)^2 = \frac{1}{6} x (1+ 2x) \label{appindistrans}
\end{equation}
\begin{equation}
\sum_{y=0}^x \frac{(x-y)y}{x+1} = \frac{x(x-1)}{6} \label{appb}
\end{equation}

\section{Coincidence rates of correlated and polarization entangled photon pairs with dark counts included} \label{appdc}

\subsection{CAR, distinguishable}

\begin{eqnarray}
R_{peak} &\simeq& \mu \alpha_s \alpha_i + (\mu \alpha_s + d_s)(\mu \alpha_i + di) \\
R_{side} &\simeq& (\mu \alpha_s + d_s)(\mu \alpha_i + d_i) \\
C_p &\simeq& 1+\frac{\mu \alpha_s \alpha_i}{(\mu \alpha_s + d_s)(\mu \alpha_i + di)}
\end{eqnarray}

\subsection{CAR, indistinguishable}

\begin{eqnarray}
R_{peak} &\simeq& \mu \alpha_s \alpha_i + \mu^2 \alpha_s \alpha_i + (\mu \alpha_s + d_s)(\mu \alpha_i + d_i) \\
R_{side} &\simeq& (\mu \alpha_s + d_s)(\mu \alpha_i + di) \\
C_t &\simeq& 1+\frac{\mu(\mu+1) \alpha_s \alpha_i}{(\mu \alpha_s + d_s)(\mu \alpha_i + di)}
\end{eqnarray}

\subsection{Entanglement, distinguishable}

\begin{eqnarray}
R_{HHp} &\simeq& \frac{1}{2}\mu \alpha_s \alpha_i + \left(\frac{1}{2}\mu \alpha_s + d_s\right)\left(\frac{1}{2}\mu \alpha_i + di\right) \\
R_{HVp} &\simeq& \left(\frac{1}{2} \mu \alpha_s + d_s\right)\left(\frac{1}{2} \mu \alpha_i + d_i\right) \\
R_{H+p} &\simeq& \frac{1}{4}\mu \alpha_s \alpha_i + \left(\frac{1}{2}\mu \alpha_s + d_s\right)\left(\frac{1}{2}\mu \alpha_i + di\right)
\end{eqnarray}

\subsection{Entanglement, indistinguishable}

\begin{eqnarray}
R_{HHindis} &\simeq& \alpha_s \alpha_i \left(\frac{\mu^2}{2} + \frac{\mu}{2}\right) + \frac{\mu \alpha_s d_i}{2} + \frac{\mu \alpha_i d_s}{2} + d_s d_i\\
R_{HVindis} &\simeq& \frac{\mu^2 \alpha \alpha_i}{4} + \frac{\mu \alpha_s d_i}{2} + \frac{\mu \alpha_i d_s}{2} + d_s d_i \\
R_{H+indis} &\simeq& \alpha_s \alpha_i \left(\frac{3\mu^2}{8} + \frac{\mu}{4}\right) + \frac{\mu \alpha_s d_i}{2} + \frac{\mu \alpha_i d_s}{2} + d_s d_i
\end{eqnarray}

\section{Coincidence rates of time-bin entanglement with dark counts included} \label{apptb}

Here, the average pair number $\mu$ is defined per two time slots. In contrast, the average pair number per pulse was used in equations in \cite{longd}. Therefore, the equations in \cite{longd} provide the same results as the equations for distinguishable pairs shown below.  

\subsection{Distinguishable}

\begin{eqnarray}
R_{aap} &\simeq& \frac{1}{8}\mu \alpha_s \alpha_i + \left(\frac{1}{4}\mu \alpha_s + d_s\right)\left(\frac{1}{4}\mu \alpha_i + di\right) \\
R_{abp} &\simeq& \left(\frac{1}{4} \mu \alpha_s + d_s\right)\left(\frac{1}{4} \mu \alpha_i + d_i\right) \\
R_{a+p} &\simeq& \frac{1}{16}\mu \alpha_s \alpha_i + \left(\frac{1}{4}\mu \alpha_s + d_s\right)\left(\frac{1}{4}\mu \alpha_i + di\right)
\end{eqnarray}

\subsection{Indistinguishable}

\begin{eqnarray}
R_{aaindis} &\simeq& \alpha_s \alpha_i \left(\frac{\mu^2}{8} + \frac{\mu}{8}\right) + \frac{\mu \alpha_s d_i}{4} + \frac{\mu \alpha_i d_s}{4} + d_s d_i\\
R_{abindis} &\simeq& \frac{\mu^2 \alpha \alpha_i}{16} + \frac{\mu \alpha_s d_i}{4} + \frac{\mu \alpha_i d_s}{4} + d_s d_i \\
R_{a+indis} &\simeq& \alpha_s \alpha_i \left(\frac{3\mu^2}{32} + \frac{\mu}{16}\right) + \frac{\mu \alpha_s d_i}{4} + \frac{\mu \alpha_i d_s}{4} + d_s d_i
\end{eqnarray}

\end{document}